\documentclass[aps,prl,twocolumn,superscriptaddress,preprintnumbers,longbibliography]{revtex4-2}
\usepackage{amsmath}
\usepackage{amssymb}
\usepackage{graphicx}
\usepackage{siunitx}

\usepackage{booktabs}

\usepackage[colorlinks = true, citecolor = blue, linkcolor = blue, urlcolor = blue]{hyperref}
\usepackage{xcolor}
\usepackage{physics}

\def\bea#1\eea{\begin{align}#1\end{align}}
\newcommand{\nn}{\nonumber\\}
\newcommand{\bef}{\begin{figure}[h!tb]\centering}
\newcommand{\eef}{\end{figure}}
\newcommand{\med}{\text{med}}
\newcommand{\nonpert}{\text{np}}

\newcommand{\bmax}{b_\text{max}}
\allowdisplaybreaks

\newcommand{\update}[1]{#1}

\begin{document}

\title{Energy--Energy Correlator for jet production in $pp$ and $pA$ collisions}
\author{Jo\~{a}o Barata}
\email{joao.lourenco.henriques.barata@cern.ch}
\affiliation{CERN, Theoretical Physics Department, CH-1211, Geneva 23, Switzerland}

\author{Zhong-Bo Kang}
\email{zkang@physics.ucla.edu}
\affiliation{Department of Physics and Astronomy, University of California, Los Angeles, CA 90095, USA}
\affiliation{Mani L. Bhaumik Institute for Theoretical Physics, University of California, Los Angeles, CA 90095, USA}
\affiliation{Center for Frontiers in Nuclear Science, Stony Brook University, Stony Brook, NY 11794, USA}

\author{Xo{\'{a}}n Mayo L\'{o}pez}
\email{xoan.mayo.lopez@usc.es}
\affiliation{Instituto Galego de F{\'{i}}sica de Altas Enerx{\'{i}}as,  Universidade de Santiago de Compostela, Santiago de Compostela 15782, Galicia, Spain}

\author{Jani Penttala}
\email{janipenttala@physics.ucla.edu}
\affiliation{Department of Physics and Astronomy, University of California, Los Angeles, CA 90095, USA}
\affiliation{Mani L. Bhaumik Institute for Theoretical Physics, University of California, Los Angeles, CA 90095, USA}

\preprint{CERN-TH-2024-201}

\begin{abstract}
In this Letter, we study the collinear limit of the Energy--Energy Correlator (EEC) in single-inclusive jet production in proton--proton ($pp$) and proton--nucleus ($pA$) collisions. We introduce a non-perturbative model that allows us to describe the EEC in the entire angular region of the current experiments. Our results for $pp$ collisions show excellent agreement with CMS and ALICE data over a wide range of jet transverse momenta. For $pA$ collisions, we include modifications from the nuclear medium, and our predictions align well with the trends observed in recent ALICE measurements.
\end{abstract}

\maketitle

\textit{Introduction.} Hadronic jets---energetic, collimated cascades of QCD particles produced in high-energy scattering experiments---are among the most powerful tools for probing the nature and properties of QCD. In events with low final-state particle multiplicity, such as those typically observed in $ee$ and $pp$ collisions, jets enable precise tests of the soft and collinear dynamics of QCD, see e.g.~\cite{Larkoski:2017jix,Salam:2010nqg,Sapeta:2015gee}. In contrast, in collisional systems such as $eA$, $pA$, and $AA$, the jet radiative cascade gets modified by the underlying QCD medium, whether it be the cold nuclear matter characterized by a highly occupied gluonic state~\cite{Gelis:2010nm,Morreale:2021pnn}, or the hot quark--gluon plasma (QGP)~\cite{Busza:2018rrf}. By comparing these modified jets to those produced in $ee$ or $pp$ collisions, one can investigate the medium-induced imprints on the jet cascade, providing a unique window into the many-body properties of QCD.

One of the key challenges in describing the vacuum fragmentation pattern of jets and extracting medium-induced modifications to jets lies in identifying observables that allow for both a theoretically consistent description and experimental accessibility across a broad kinematic region. Recently, it has been argued that correlations between asymptotic fluxes of certain currents
can satisfy these requirements, see e.g.~\cite{Craft:2022kdo,Lee:2022ige,Dixon:2019uzg,Moult:2018jzp,Kang:2024dja,Liu:2022wop,Chen:2020adz,Lee:2023npz,CMS:2024ovv,CMS:2024mlf,ALICE:2024dfl}. Focusing on the widely studied case of Energy-Energy Correlators (EECs)~\cite{Basham:1979gh,Basham:1978bw,Hofman:2008ar,Sveshnikov:1995vi}, in vacuum they allow, for example, to directly access the anomalous scaling of leading-twist QCD operators at high precision~\cite{Hofman:2008ar,Dixon:2019uzg,Lee:2022ige,Komiske:2022enw} and cleanly separate the relevant scales inside jets, 
sharply splitting perturbatively and non-perturbatively dominated sectors~\cite{Lee:2022ige,Komiske:2022enw,ALICE:2024dfl}. Extending this program to $eA$, $pA$, and $AA$ collisions,
EECs have the potential to directly probe the scales of the underlying matter and resolve its spatial structure~\cite{Singh:2024vwb,Barata:2023bhh,Yang:2023dwc,Xing:2024yrb,Barata:2023zqg, Devereaux:2023vjz, Li:2020bub,Li:2021txc,Liu:2022wop,Liu:2023aqb,Cao:2023oef,Li:2023gkh,Bossi:2024qho,Fu:2024pic}.

Despite this promise, the theoretical control of EECs in complex collisional systems, particularly $AA$ collisions, remains limited. Current studies are mostly constrained to (partial) leading-order perturbative calculations~\cite{Barata:2023bhh,Yang:2023dwc,Andres:2022ovj} and numerical model simulations~\cite{Bossi:2024qho,Yang:2023dwc}. Even in cleaner environments such as $ep$ or $pp$ collisions, the description of EECs in the non-perturbative sector is still poorly understood.

This non-perturbative region of EECs has recently garnered a lot of interest due to the emergent scaling behavior, indicative of the hadronization transition's universality~\cite{CMS:2024mlf,CMS:2024ovv,ALICE:2024dfl,Komiske:2022enw,Jaarsma:2023ell}.
In the back-to-back limit of EECs where the transverse-momentum-dependent (TMD) factorization is valid, the non-perturbative region can be described using non-perturbative Sudakov factors, related to the transverse momentum dependence of fragmentation, and the non-perturbative Collins--Soper (CS) kernel required for a resummation of large logarithms in rapidity~\cite{Collins:1981zc,deFlorian:2004mp,Tulipant:2017ybb,Kardos:2018kqj,Kang:2024dja}.
In the collinear limit, the CS resummation is not needed, but the physics related to the non-perturbative fragmentation is expected to remain the same, see e.g.~\cite{Liu:2024lxy}.
Exploring the universality of these non-perturbative effects is then of high interest due to its potential in describing a vast range of observables across different experimental environments.

To develop a comprehensive understanding of jet substructure and medium-induced modifications from EECs, it is essential to have theoretical predictions that encompass both the perturbative and the non-perturbative sectors of $pp$ collisions, while also incorporating medium-induced modifications for systems interpolating between $pp$ and $AA$ events.
A prime candidate for this is $pA$ collisions where high-energy jets can still be produced and fragment in the presence of cold nuclear matter~\cite{Liu:2022ijp,Albacete:2010bs,Levin:2010dw,Stasto:2014sea,Blaizot:2004wu,Blaizot:2004wv,Bury:2017xwd,Dumitru:2005gt,Liu:2019iml}. This constitutes a \textit{cleaner} environment when compared to the QGP, allowing us to directly probe the medium-induced modifications to EECs inside jets.

In this Letter, we present a formalism to describe the collinear limit of the EEC distribution in $pp$ and $pA$ collisions. Our framework accounts, at the same time, for both the non-perturbative regime, where a parametrization inspired by TMD factorization is implemented~\cite{Kang:2024dja,Guo:2024vpe}, and the perturbative sector, where we employ the leading-logarithmic (LL) evolution of the vacuum cascade as well as the medium-induced modifications to the partonic branchings in $pA$ events. This holistic approach allows for an efficient description of the entire EEC region and can be extended to other collisional systems. As we shall show, this permits a first direct observation of purely medium-induced modifications to the jet cascade in $pA$ collisions using EECs.

\textit{Proton--proton collisions.} 
The collinear limit of the EEC distribution satisfies the factorization formula~\cite{Dixon:2019uzg,Lee:2022ige}:
\begin{align}\label{eq:fact_th}
 \frac{\dd \Sigma^{pp}}{\dd p_T\, \dd y \,\dd R_L}  &= \int_0^1 \dd x \, x^2 \frac{\dd J\left(x,p_T ,R_L\right)}{\dd R_L}  \vdot  H\left(x,y,p_T\right) \, ,
\end{align}
where we made implicit the dependence on the factorization scale $\mu$, $y$ denotes the rapidity, and $p_T$ is the transverse momentum of the jet. 
This factorization formula separates the production of the outgoing states into a hard function $H$, which does not depend on the angular separation between the outgoing states $R_L$, and an EEC jet function $J$, which incorporates the collinear fragmentation inside the jet. Restraining our discussion to the LL accuracy, the convolution in Eq.~\eqref{eq:fact_th} trivializes to setting $x=1$, and the dot product is taken in flavor space, i.e. $J=\{J_q,J_g\}$, and respectively for $H$.

The hard function can be obtained from collinear factorization~\cite{Collins:1989gx,Bodwin:1984hc,Collins:1988ig,Collins:2011zzd}, where the cross section to extract a parton $c$ from the initial hard scattering reads
\begin{align}\label{eq:PDF}
\frac{\dd \sigma_c}{\dd{p_T}\,\dd y} =& \sum_{a,b}f_{a/p}(x_a, \mu) \otimes f_{b/p}(x_b, \mu) \otimes  \hat \sigma_{a+b\to c} \, .
\end{align}
Here $\otimes$ denotes a convolution over the parton momentum fraction, $\hat\sigma_{ab\to c}$ is the hard partonic cross section to produce a parton $c$ with a transverse momentum $p_T$ and rapidity $y$, and $f_{a/p}$ is the parton distribution function (PDF) associated to the parton $a$ coming from the proton. The hard function can then be defined from Eq.~\eqref{eq:PDF} as $H(y,p_T,\mu) = \{\dd\sigma_q/(\sigma\,\dd p_T\,\dd y), \dd\sigma_g/(\sigma\,\dd p_T\, \dd y)\}$~\cite{Owens:1986mp,Kang:2016mcy,Kang:2018vgn,Lee:2024icn}, which is normalized by the jet cross section $\sigma$ as in the definition of the EEC. \update{We choose the factorization scale $\mu = p_T$ for the central predictions shown in the plots below.}

Moving to the jet function, we first note that renormalization group (RG) consistency implies that, at the LL accuracy, $J$ satisfies a multiplicative renormalization formula, which results in the differential jet function~\cite{Dixon:2019uzg}
\begin{align}\label{eq:J_vac}
\frac{\dd J }{\dd R_L} &= \frac{\alpha_s(R_L p_T)}{\pi R_L }\,  (1,1) \vdot
\left(\frac{\alpha_s(\mu)}{\alpha_s(R_L p_T)}\right)^{\frac{\gamma(3)}{\beta_0}}  \vdot   \gamma(3)\, ,
\end{align}
where $\beta_0=(11C_A-2n_f)/3$ and $\gamma(j)\equiv -\int_0^1 \dd z \, z^{j-1} \hat P(z)$ is the anomalous dimension matrix, 
with $\hat P$ being the regularized splitting function, such that  only the spin $j=3$ contributes to the EEC at the leading power. As a result, Eq.~\eqref{eq:J_vac} defines the perturbative vacuum component of the EEC jet function.

So far, we have discussed the perturbative sector where $R_L p_T\gg \Lambda_{\rm QCD}$. At lower scales, the EEC becomes highly sensitive to non-perturbative dynamics in the fragmentation process.
The underlying physics of fragmentation leading to the modification of the momentum distribution should be the same as in the back-to-back limit, and thus we model it using a form inspired by the non-perturbative Sudakov factor in TMD factorization~\cite{Kang:2024dja,Liu:2024lxy}:
\begin{equation}\label{eq:J_np_pp}
    j_\nonpert(b) =  \exp(- a_0 b)  \, ,
\end{equation}
where $a_0$ is a \textit{free} parameter accounting for non-perturbative effects.
We note that the parameter $a_0$ should differ for quarks and gluons due to their different color charge, but in this work we shall not consider the separation in flavor.
In the momentum space, we can interpret $j_\nonpert$ as a probability distribution for the produced hadrons to get additional transverse momentum in the fragmentation, and the condition $j_\nonpert(b=0)=1$ guarantees that the distribution is correctly normalized. 
Furthermore, we note that this non-perturbative element inherits the universal behavior of the hadronization transition and can thus be applied to extractions of the (vacuum) jet EEC in other processes; we discuss the medium-induced modifications below. To incorporate this non-perturbative element, it is convenient to work in the position space where the convolution is multiplicative:
\begin{align}\label{eq:EEC_pos}
    \frac{\dd \Sigma^{pp}}{\dd p_T\, \dd y \, \dd R_L}
    &=   R_L p_T^2
    \int_0^\infty  \dd{b} b
    \,
    J_0( R_L p_T  b ) 
    \,
    j_\nonpert(b)
    \,
  \widetilde \Sigma(b) \, , \\
   \widetilde \Sigma(b)
    &= 
    (1,1) \vdot 
    \left(\frac{\alpha_s(R p_T)}{\alpha_s(\mu_{b_*})}\right)^{\frac{\gamma(3)}{\beta_0}}
    \vdot H \, .
    \label{eq:Sigma_tilde}
\end{align}
This relation follows from evolving the jet function in the position space where the initial scale $R_L p_T$ is replaced by $\mu_{b_*}= 2 e^{-\gamma_E} / b_*$, see the supplementary material. This choice minimizes large logarithms that appear at higher orders, and we use the standard $b_*$-prescription $ b_* = b/\sqrt{1 + b^2 / \bmax^2}$
with $\bmax = 2e^{-\gamma_E} \, \si{GeV^{-1}}$~\cite{Collins:1984kg,Collins:2014jpa}.
Eq.~\eqref{eq:EEC_pos} resums the same logarithms as the momentum-space expression in Eq.~\eqref{eq:J_vac}. 
We take into account the finite jet radius $R=0.4$, consistent with the experimental data, by evolving the jet function up to  $R p_T$~\cite{Kang:2017glf}.

\begin{figure*}[t!]
        \centering
        \includegraphics[width=.95\textwidth]{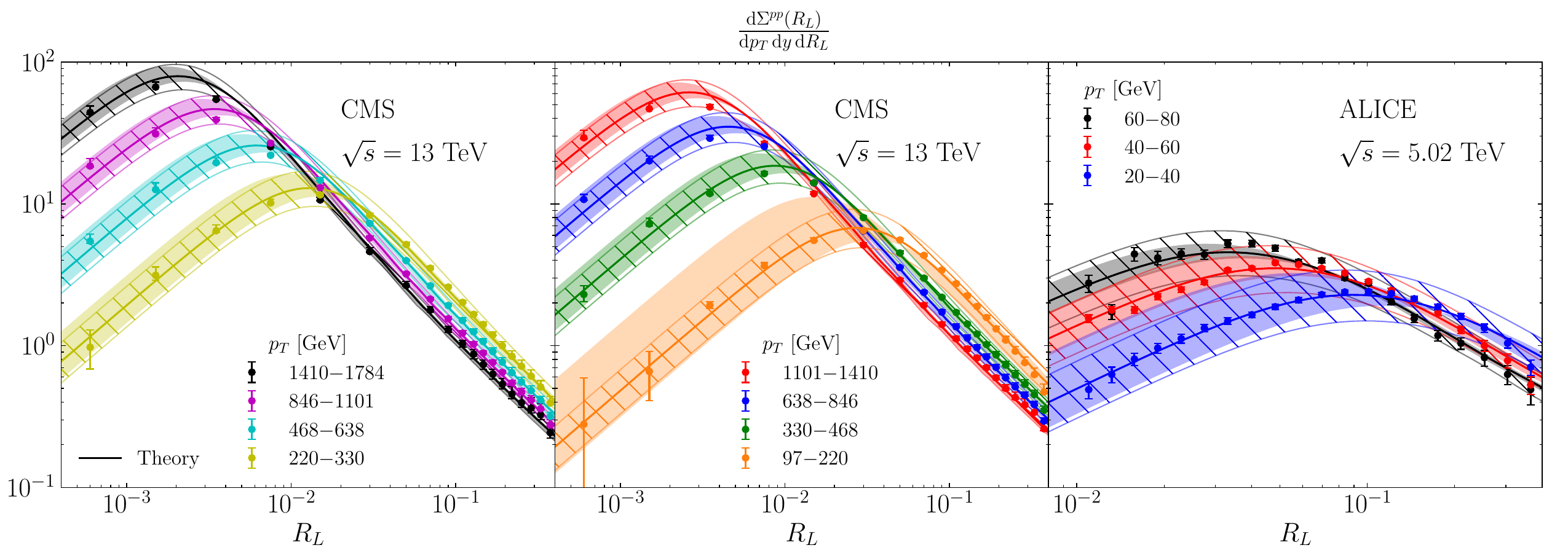}
    \caption{
    Differential EEC distribution in $pp$ collisions as a function of the angle $R_L$ compared to the experimental data from CMS~\cite{CMS:2024mlf} and ALICE~\cite{ALICE:2024dfl}.
    The theory prediction at the average $p_T$ of the bin is shown with solid lines, while the shaded band corresponds to the prediction with the minimum and maximum $p_T$ in each bin. 
    \update{The hatched bands indicate the theoretical uncertainty associated with resummation scale variation, obtained by independently varying the scales $\mu_{b_\star}$ in Eq.~\eqref{eq:Sigma_tilde} and $\mu$ in Eq.~\eqref{eq:PDF} up and down by a factor of 2, and taking the envelope of the resulting predictions.}
    }
    \label{fig:pp}
\end{figure*}

The numerical results for the EEC in $pp$ collisions are shown in Fig.~\ref{fig:pp}, with the parameter $a_0$ characterizing non-perturbative physics chosen as
$a_0 = \SI{3.8}{GeV}$ for CMS and
$a_0 = \SI{2.5}{GeV} $ for ALICE, fitted to the experimental data presented in \cite{CMS:2024mlf} and \cite{ALICE:2024dfl}, respectively.
We use the CT18ANLO fit \cite{Hou:2019qau} for the proton PDF, and we also incorporate a $K$-factor of $K \approx 0.90$ for CMS and $K \approx 0.75$ for ALICE to account for higher-order corrections beyond the LL approximation in our formalism.
\update{The difference in $a_0$ arises from the experimental definitions of jet $p_T$~\cite{Jacak_discussion}: ALICE reports charged jets, where $p_T$ is the sum over charged particles only, while CMS uses inclusive jets, summing over both charged and neutral particles. Assuming isospin symmetry, rescaling the ALICE value by a factor of $3/2$ yields $a_0 \approx \SI{3.75}{GeV}$, consistent with the CMS value. This consistency highlights the robustness of our model across different jet definitions and kinematic regimes.}

A direct analysis of Fig.~\ref{fig:pp} allows us to conclude that our description, accommodating the LL perturbative prediction at large angles and a non-perturbative jet function at small angles, can describe the data quite precisely in the entire angular domain and for a 
wide range of different $p_T$ values. We note that while the large-angle region is fully determined from perturbative considerations, the non-perturbative sector for all of the curves in each experiment is described by the same $a_0$ parameter, in agreement with the universal character of $j_{\rm np}$. Most crucially, we capture the location of the distribution's peak for all values of $p_T$, indicating a good theoretical description of the transition between the two sectors.

\textit{Proton--nucleus collisions.} Having introduced a description of the EEC in $pp$ collisions in the full angular domain, we extend our construction to collisional systems where a QCD background medium is present. 
The presence of the medium can, in general, induce modifications to both the hard and the EEC jet functions. In $pA$ collisions, the hard function should be updated to account for the target nucleus. To that end we replace one of the PDFs in Eq.~\eqref{eq:PDF} by a nuclear PDF (nPDF), i.e. $f_{b/p}(x_b, \mu)\to f_{b/A}(x_b, \mu)$. 

The jet function receives medium-induced modifications to the perturbative and non-perturbative sectors. For the former, the medium correction enters as a higher-twist/power correction to the leading perturbative QCD result. Thus, it cannot change the RG evolution at the current accuracy, and the corrections can be entirely incorporated into the fixed-order jet function which now reads $J^{pA} = J^{pp} + J^{\rm med}$, with
\begin{align}
\frac{\dd J^{\rm med}}{\dd R_L} &= \frac{\alpha_s (R_L p_T)}{\pi R_L}\nn
&\times \int_0^1 \dd x \, x(1-x) P^{\rm vac}(x)  F_{\rm med}(p_T,R_L,x) \, ,
\end{align}
where the medium modification (matrix) factor $F_{\med}$ depends on  the jet quenching parameter $\hat q$ and the effective path length in the medium $L$. Here $P^{\rm vac}$ denotes the appropriate vacuum splitting kernel. 
In the large-$N_c$ limit, $F_{\rm med}$ can be explicitly computed in a semi-classical expansion~\cite{Isaksen:2020npj, Isaksen:2023nlr}; we provide the explicit formulas in the supplementary material.

The non-perturbative sector of $J^{pA}$ is also modified compared to $pp$ collisions. Physically, these corrections are driven by the multiple scattering of the jet partons on the nucleus, which results in the broadening of the underlying momentum distribution.
Thus, at small angles we update Eq.~\eqref{eq:J_np_pp} to the form 
\begin{equation}\label{eq:J_np_pA}
    j_\nonpert(b) =  \exp(- a_0 b - a_1 b^2)  \, ,
\end{equation}
where $a_0$ is the same as in the vacuum case and the parameter $a_1$ accounts for the medium-induced diffusive contribution to the hadronization of the jet partons. Note that $a_1$ 
captures the relative transverse momentum broadening of a particle pair inside the jet, differing, in general, from the standard $\hat q L$ scaling for a single parton. 
Finally, we also note that, as in $AA$ collisions, one should account for energy-loss effects which can affect the non-perturbative sector. 
However, since in $pA$ collisions these are expected to be negligible due to the smallness of $\hat q$ and $L$, we shall neglect them in what follows.

We consider a typical value for the jet quenching parameter in cold nuclear matter, $\hat q=\SI{0.02}{GeV^2/fm}$~\cite{Zhang:2021tcc,Ru:2019qvz}, and take the path length to be $L=\SI{3}{fm}$, while the non-perturbative medium-dependent parameter is set at $a_1=\SI{0.25}{GeV^2}$. 
The $K$-factors, the proton PDF and the $a_0$ parameter are the same as in our $pp$ analysis, while we use the EPPS21 fit~\cite{Eskola:2021nhw} for the nPDF.

\begin{figure}[t!]
        \centering
        \includegraphics[width=.4\textwidth]{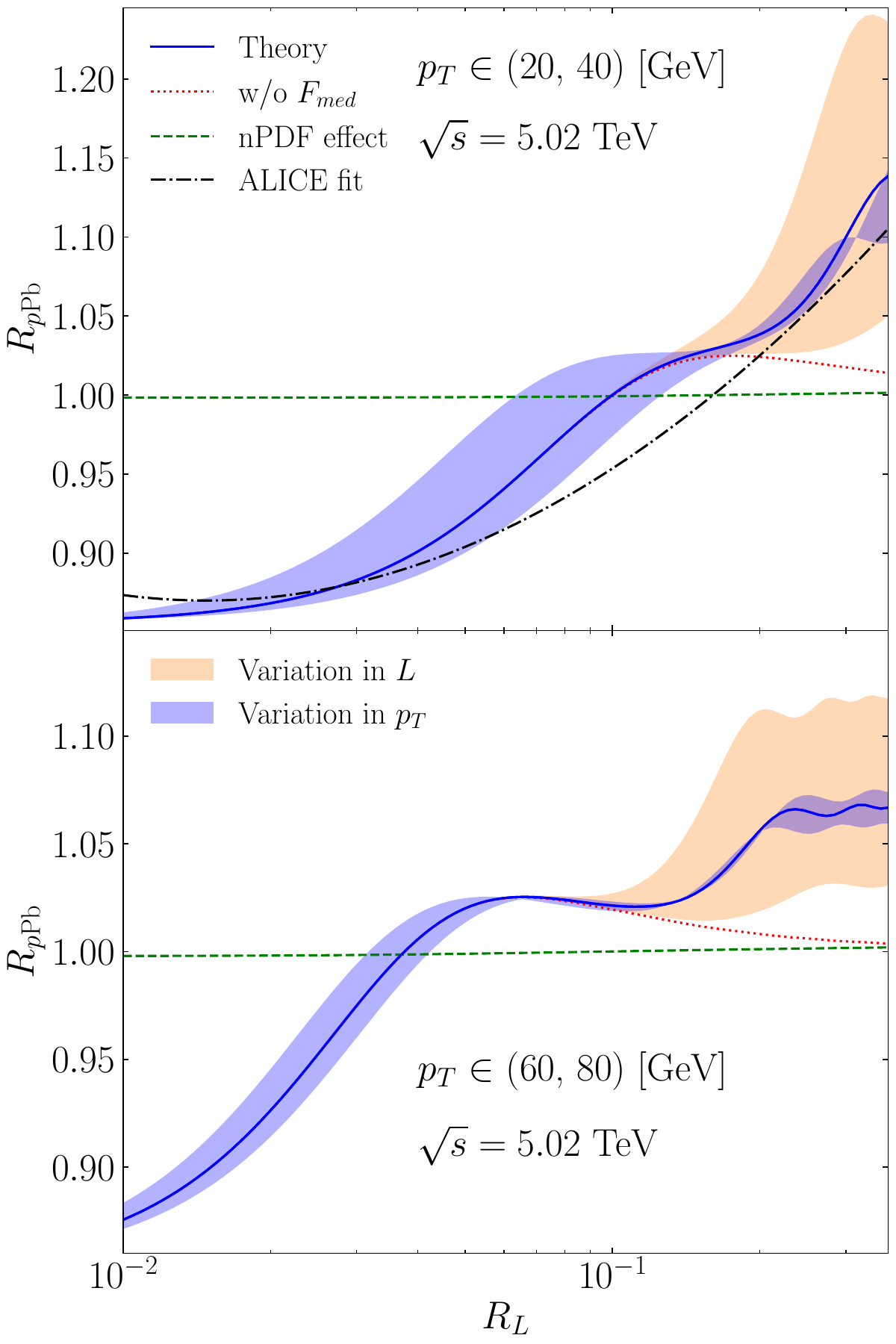}
    \caption{
    \update{
    Ratio of the differential EEC $p \rm Pb$ distribution with respect to $pp$ collisions, as a function of $R_L$, compared to the experimental fit provided by ALICE~\cite{ALICEpA} (dot-dashed black line):
    $R^{\rm ALICE}_{p \rm Pb} = 0.87 + 0.121(\log_{10}(R_L) + 1.831)^2$.
    The theory prediction, at the average $p_T$ of the bin, is shown by the blue solid line. The ratio with $F_{\rm med}= 0$ (dotted red line), and the sole effect of introducing the nPDF in the vacuum distribution (dashed green line) are also shown. The blue shaded band corresponds to the prediction using the minimum and maximum $p_T$ of the bin, while the orange band corresponds to modifying the length of the medium $L\in (2,4)~$fm.} 
    }
    \label{fig:pPb-pp ratio}
\end{figure}

The numerical results for the ratio between the differential EEC in $p \rm Pb$ with respect to $pp$ collisions as a function of the opening angle, defined as $R_{pA} = \frac{\dd\Sigma^{pA}(R_L)}{\dd p_T \, \dd y \, \dd R_L} / \frac{\dd\Sigma^{pp}(R_L)}{\dd p_T \, \dd y \, \dd R_L}$, are shown in Fig.~\ref{fig:pPb-pp ratio}, along with the fit to the experimental data in the $p_T\in (20,40)$ GeV bin provided by the ALICE collaboration~\cite{ALICEpA} \update{(dot-dashed black line)}. We first note that the effect of the nPDF \update{(dashed green line)} is almost negligible for the whole $R_L$ range. This is expected since the EEC is normalized by the jet cross section, and the effect of nPDF mostly cancels out in the ratio. We then present two predictions, one including only the medium modification to the non-perturbative sector \update{(dotted red line)} and one including all the medium effects (solid blue line). As is evident when comparing to the experimental fit, our modification to $j_{\rm np}$ is crucial to capture the data trend for a suppression in $R_{pA}$. This gives a clear indication of the importance of momentum broadening in the nucleus at small transverse momentum scales. We note that, so far and to our best knowledge, momentum diffusion had only been taken into account in the perturbative region of the EEC. 

Comparing the experimental fit to the theory at large $R_L$, we clearly observe the enhancement effect driven by the medium modification factor $F_{\rm med}$. Note that at small angles $F_{\rm med}\approx 0$, and thus it does not affect the non-perturbative region. Moreover, compared to the case of $AA$ collisions 
where there is a competing effect driven by the soft physics of the bulk~\cite{Yang:2023dwc,BarataMedRes,Bossi:2024qho}, in the $pA$ scenario the medium does not significantly contaminate the jet. 
\update{As a result,
our analysis provides first evidence for the emergence of medium-induced jet modifications in the EEC distribution, within a simple and improvable theoretical model.}
We further note that the approximations used to compute $F_{\rm med}$ in this work tend to overestimate the enhancement, and thus improved theoretical descriptions would most likely drive the model curve closer to the experimental fit, see discussion in~\cite{Barata:2023bhh}. We also provide a prediction for jets with  $p_T \in (60, 80)$ GeV in the lower panel of Fig.~\ref{fig:pPb-pp ratio} to illustrate the $p_T$-dependence of the nuclear modification.

{\it Conclusions.} In this Letter, we have presented a theoretical framework for the differential EEC distribution in the collinear limit, encompassing both $pp$ and $pA$ collisions. Our model incorporates the leading-logarithmic form of the observable in the perturbative sector and employs a TMD-inspired parametrization to describe the non-perturbative regime. By fitting the available CMS and ALICE $pp$ data, we have achieved excellent agreement with experiment, successfully describing the EEC at both small and large angles and accurately predicting the location of the distribution's peak. For $pA$ collisions, our formalism incorporates nuclear medium effects, enabling a unified description across both the non-perturbative and the perturbative regions.

In the small-angle region, we have highlighted the critical role played by medium-induced broadening, essential to accurately capture the hadronization transition. At large angles, we have included the leading medium-induced power corrections, demonstrating that these lead to a significant enhancement compared to $pp$ data. Contrasting our theoretical predictions with ALICE's fit to experimental data~\cite{ALICEpA}, we have shown that our framework provides a precise and comprehensive description of the measurements.

This work establishes a solid foundation for a future detailed and thorough understanding of the jet EEC across a diverse set of experiments, opening exciting opportunities for future research.

\vspace{0.2 cm}
We thank B.~Jacak and A.~Nambrath for their valuable discussions regarding the ALICE measurements. X.M.L. thanks the nuclear physics group at UCLA for their hospitality during the development of this work. Z.K. and J.P. are supported by the National Science Foundation under grant No.~PHY-1945471. The work of X.M.L is supported by European Research Council project ERC-2018-ADG-835105 YoctoLHC; by Xunta de Galicia (Centro singular de investigación de Galicia accreditation 2019-2022), by European Union ERDF; and by Grant CEX2023-001318-M funded by MICIU/AEI/10.13039/501100011033 and by ERDF/EU. X.M.L.'s contribution to this work is also supported under scholarship No.~PRE2021-097748, funded by MCIN/AEI/10.13039/501100011033 and FSE+. 
This work is also supported by the U.S. Department of Energy, Office of Science, Office of Nuclear Physics, within the framework of the Saturated Glue (SURGE) Topical Theory Collaboration.

\bibliographystyle{JHEP-2modlong.bst}

\bibliography{refs.bib}

\clearpage
\onecolumngrid

\section*{Supplementary material}

\subsection{Evolution of the jet function in position space}

In this appendix, we describe the evolution of the jet in the position space instead of the momentum space.
We define the position-space jet function $j(b)$ as the Fourier transform of the differential jet function $\dd J / \dd[2]{\vec k_T}$,
\begin{equation}
 j(b) = \int \dd[2]{\vec k_T} 
     \frac{\dd{J}}{ \dd[2]{\vec k_T}} e^{i \vec b \vdot \vec k_T},
\end{equation}
where $k_T = p_T R_L$ and
\begin{equation}
     \frac{\dd{J}}{ \dd[2]{\vec k_T}}
     = \frac{1}{ 2\pi R_L p_T^2 } \frac{\dd{J}}{ \dd{R_L}}= \delta^2(\vec k_T) \cdot  (1,1)
\end{equation}
at leading order. 
This corresponds to $j(b) = (1,1)$ in the position space.

The evolution equation for the jet function resumming the large logarithms is not modified when going from the momentum to the position space, and it reads:
\begin{equation}
\label{eq:jet_evolution}
    \frac{ \dd j(b,\mu)}{\dd \ln\mu^2} = - \frac{\alpha_s(\mu)}{4\pi} j(b,\mu) \vdot \gamma(3) \, .
\end{equation}
Keeping only the LO $\beta$-function in the running of the coupling,
\begin{equation}
    \frac{\dd \alpha_s(\mu)}{\dd \ln \mu^2} = - \frac{1}{4\pi} \alpha_s^2(\mu) \beta_0,
\end{equation}
the solution to Eq.~\eqref{eq:jet_evolution} can be written as
\begin{equation}
    j(b, \mu) = j(b, \mu_0) \vdot \qty( \frac{\alpha_s(\mu)}{\alpha_s(\mu_0)} )^{\frac{\gamma(3)}{\beta_0}}.
\end{equation}
Evolving to the final scale $\mu = R p_T$ and taking the initial condition as $j(b, \mu_{b_*}) = (1,1)$,
we then get 
\begin{equation}
    j(b, R p_T) = (1,1) \vdot \qty( \frac{\alpha_s(R p_T)}{\alpha_s(\mu_{b_*})} )^{\frac{\gamma(3)}{\beta_0}}.
\end{equation}
This can be combined with the hard function to define the position-space EEC
\begin{equation}
    \widetilde \Sigma(b) = j(b, R p_T) \vdot H,
\end{equation}
which 
corresponds to 
Eq.~\eqref{eq:Sigma_tilde} in the main article.

\subsection{Medium modification factors}
In this appendix, we provide the explicit form of both the vacuum splitting functions and  the modification factors due to the propagation inside a medium of length $L$ for all the relevant channels:  $q\to qg$, $g\to gg$ and $g\to q \bar{q}$. The unregularized vacuum splitting functions read
\begin{align}
    &P^{\rm vac}_{qq}(x) = \, C_F \frac{1+x^2}{1-x} \, , \qquad \qquad \qquad \qquad \quad \quad \; P^{\rm vac}_{gq}(x)=P^{\rm vac}_{qq}(1-x) \, , \notag
    \\ & P^{\rm vac}_{qg}(x) = 2 \, n_f \, T_R \left(x^2+(1-x)^2\right)\, , \qquad \qquad \quad P^{\rm vac}_{gg}(x)=2 \, C_A \left(\frac{x}{1-x}+\frac{1-x}{x} + x (1-x)\right) \, .
\end{align}
By following the procedure developed in~\cite{Isaksen:2020npj,Isaksen:2023nlr}, see also~\cite{Apolinario:2014csa,Kovner:2001vi}, we obtain the medium modifications factors. For any flavor channel, they can be written as 
\begin{align}\label{eq:nice_F}
    F_{\text{med}}(x,R_L) = 2 \int_0^L \frac{\dd t_1}{t_f} \left[\int_{t_1}^L \frac{\dd t_2}{t_f} \cos\left(\frac{t_2-t_1}{t_f}\right) \mathcal{C}^{(4)}(L,t_2) \mathcal{C}^{(3)}(t_2,t_1)-\sin\left(\frac{L-t_1}{t_f}\right)\mathcal{C}^{(3)}(L,t_1)\right] \, ,
\end{align}
with $\mathcal{C}^{(n)}(t_f,t_i)$ the $n$-particle correlator of light-like Wilson lines in the light-cone time interval $(t_f,t_i)$, which can be computed in the large-$N_c$ limit by using the harmonic oscillator approximation. Note that to obtain Eq.~\eqref{eq:nice_F} we have resorted to the so-called \textit{tilted} Wilson line approximation, see e.g.~\cite{Altinoluk:2014oxa}.

\begin{enumerate}
    \item {$q\to qg$:} The quark--gluon splitting modification factor is deduced in the aforementioned works and reads
\begin{align}
    F^{qq}_{\text{med}}(x,R_L) = 2 \int_0^L\frac{\dd t_1}{t_f} \bigg[&\int_{t_1}^L \frac{\dd t_2}{t_f} \, \cos\left(\frac{t_2-t_1}{t_f}\right) \, e^{-\frac{\hat{q}}{12} R_L^2 (t_2-t_1)^3 \left(1+(1-x)^2\right)}\, e^{-\frac{\hat{q}}{4} R_L^2 (L-t_2) (t_2-t_1)^2 \left(1-2(1-x)+3(1-x)^2\right)} \nn 
    & \times \left(1-\frac{\hat{q}}{2}R_L^2 x (1-x) (t_2-t_1)^2 \int_{t_2}^L\dd s \, e^{-\frac{\hat{q}}{12} R_L^2 \left((s-t_2)^2 (2s-3t_1+t_2) + 6x(1-x)(s-t_2)(t_2-t_2)^2\right)}\right) \nn 
    &- \sin\left(\frac{L-t_1}{t_f}\right) \, e^{-\frac{\hat{q}}{12}R_L^2 (L-t_1)^3 \left(1+(1-x)^2\right)}  \bigg]\,,
\end{align}
with $R_L$ being the angle between the two prongs, $x$ their energy fraction, $p_T$ the transverse momentum of the jet, and the formation time defined as $t_f = \frac{2}{x(1-x) R_L^2 p_T}$. 

\item {$g\to gg$:} For this channel, the three-particle correlator is also directly given in the above works:
\begin{align}
    \mathcal{C}^{(3)}_{g\to gg}(t_2,t_1) = e^{-\frac{\hat{q}}{6} R_L^2 (t_2-t_1)^3 (1-x+x^2)} \, .
\end{align}
In order to obtain $\mathcal{C}^{(4)}$ for the gluon splitting, the differential equation Eq.~(4.12) in \cite{Isaksen:2020npj} must be solved. Doing so, one gets
\begin{align}
    \mathcal{C}^{(4)}_{g\to gg}(L,t_2) &= e^{-\frac{\hat{q}}{4}R_L^2 (L-t_2) (t_2-t_1)^2 (x^2+(1-x)^2)} \, e^{-\frac{\hat{q}}{12} R_L^2 \left((L-t_1)^3 + (L-t_2)^3 - (t_2-t_1)^3\right)} \notag
    \\ &\times \left(1+\frac{\hat{q}}{2} R_L^2 \int_{t_2}^L\dd s \, (s-t_1) (s-t_2) \, e^{\frac{\hat{q}}{12} R_L^2 \left((s-t_1)^2+(s-t_2)^3-(t_2-t_1)^3\right)} \, e^{-\frac{\hat{q}}{4} R_L^2 (s-t_2) (t_2-t_1)^2 \left(x^2+(1-x)^2\right)}\right) \, .
\end{align}
Thus, the medium modification factor for the gluon splitting channel can be written as 
\begin{align}
    F^{gg}_{\text{med}}(x,R_L) &= 2\int_0^L \frac{\dd t_1}{t_f}\bigg[\int_{t_1}^L \frac{\dd t_2}{t_f} \, \cos\left(\frac{t_2-t_1}{t_f}\right) \, e^{-\frac{\hat{q}}{4}R_L^2 (L-t_2) (t_2-t_1)^2 (x^2+(1-x)^2)} \,  \notag
    \\ & \times e^{-\frac{\hat{q}}{12} R_L^2 \left((L-t_1)^3 + (L-t_2)^3 - (t_2-t_1)^3\right)} \, e^{-\frac{\hat{q}}{6} R_L^2 (t_2-t_1)^3 (1-x+x^2)} \, \notag 
    \\ & \times \bigg(1 +\frac{\hat{q}}{2} R_L^2 \int_{t_2}^L\dd s \, (s-t_1) (s-t_2) \, e^{\frac{\hat{q}}{12} R_L^2 \left((s-t_1)^2+(s-t_2)^3-(t_2-t_1)^3\right)} \, e^{-\frac{\hat{q}}{4} R_L^2 (s-t_2) (t_2-t_1)^2 \left(x^2+(1-x)^2\right)}\bigg) \notag
    \\ & \hspace{3cm} - \sin\left(\frac{L-t_1}{t_f}\right)\, e^{-\frac{\hat{q}}{6} R_L^2 (L-t_1)^3 (1-x+x^2)}\bigg] \,.
\end{align}
\item {$g\to q\bar q$:} Finally, the medium correction to the pair production channel can be easily computed in the corresponding limits. The three-particle correlator reads
\begin{equation}
    \mathcal{C}^{(3)}_{g\to q\bar{q}}(t_2,t_1) = e^{-\frac{\hat{q}}{12} R_L^2 (t_2-t_1)^3 \left(x^2 + (1-x)^2\right)} \, ,
\end{equation}
while the four-particle correlator reads 
\begin{align}
    \mathcal{C}^{(4)}_{g\to q\bar{q}}(L,t_2) = e^{-\frac{\hat{q}}{4} R_L^2 (L-t_2) (t_2-t_1)^2 \left(x^2 + (1-x)^2\right)} \, . 
\end{align}
We have noticed a misprint in the equation providing $\mathcal{C}^{(4)}$ in terms of correlators of Wilson lines in~\cite{Isaksen:2020npj} for this channel, where the quadrupole and the double dipole are switched, c.f.~\cite{Barata:2024bqp,Attems:2022ubu}.
With this, the medium modification factor becomes
\begin{align}
    F^{qg}_{\text{med}} (x,R_L) = 2\int_0^L \frac{\dd t_1}{t_f} &\bigg[\int_{t_1}^L \frac{\dd t_2}{t_f} \, \cos\left(\frac{t_2-t_1}{t_f}\right) \,e^{-\frac{\hat{q}}{4} R_L^2 (L-t_2) (t_2-t_1)^2 \left(x^2 + (1-x)^2\right)} \, e^{-\frac{\hat{q}}{12} R_L^2 (t_2-t_1)^3 \left(x^2 + (1-x)^2\right)}\notag
    \\ & \hspace{0.5cm} - \sin\left(\frac{L-t_1}{t_f}\right) \, e^{-\frac{\hat{q}}{12} R_L^2 (L-t_1)^3 \left(x^2 + (1-x)^2\right)}   \bigg]\,.
\end{align}
\end{enumerate}

\end{document}